\documentclass[conference]{IEEEtran}
\IEEEoverridecommandlockouts
\usepackage{cite}
\usepackage{amsmath,amssymb,amsfonts}
\usepackage{algorithmic}
\usepackage{graphicx}
\usepackage{textcomp}
\usepackage{xcolor}
\usepackage{mathtools}
 \usepackage{nicematrix}
\usepackage[colorlinks=true, allcolors=blue, pdfborderstyle={/S/U/W 1}]{hyperref}

\newcommand{\blockcomment}[1]{}

\def\BibTeX{{\rm B\kern-.05em{\sc i\kern-.025em b}\kern-.08em
    T\kern-.1667em\lower.7ex\hbox{E}\kern-.125emX}}
\begin{document}

\title{Assessing the performance of CT image denoisers using Laguerre-Gauss Channelized Hotelling Observer for lesion detection
\blockcomment{*\\
{\footnotesize \textsuperscript{*}Note: Sub-titles are not captured in Xplore and
should not be used}
\thanks{Identify applicable funding agency here. If none, delete this.}
}}

\author{\IEEEauthorblockN{Prabhat Kc}
\IEEEauthorblockA{\textit{DIDSR/OSEL/CDRH} \\
\textit{US Food and Drug Administration}\\
Silver Spring, USA \\
\href{mailto:prabhat.kc@fda.hhs.gov}{prabhat.kc@fda.hhs.gov}}
\and
\IEEEauthorblockN{Rongping Zeng}
\IEEEauthorblockA{\textit{DIDSR/OSEL/CDRH} \\
\textit{US Food and Drug Administration}\\
Silver Spring, USA \\
\href{mailto:rongping.zeng@fda.hhs.gov}{rongping.zeng@fda.hhs.gov}}
}

\vspace{-5.5em}
\maketitle

\vspace{-4.0em}
\begin{abstract}
The remarkable success of deep learning methods in solving computer vision problems, such as image classification, object detection, scene understanding, image segmentation, etc., has paved the way for their application in biomedical imaging. One such application is in the field of CT image denoising, whereby deep learning methods are proposed to recover denoised images from noisy images acquired at low radiation. Outputs derived from applying deep learning denoising algorithms may appear clean and visually pleasing; however, the underlying diagnostic image quality may not be on par with their normal-dose CT counterparts. In this work, we assessed the image quality of deep learning denoising algorithms by making use of visual perception- and data fidelity-based task-agnostic metrics (like the PSNR and the SSIM) - commonly used in the computer vision - and a task-based detectability assessment (the LCD) – extensively used in the CT imaging. When compared against normal-dose CT images, the deep learning denoisers outperformed low-dose CT based on metrics like the PSNR (by $\textbf{2.4}$ to $\textbf{3.8}$ dB) and SSIM (by $\textbf{0.05}$ to $\textbf{0.11}$). However, based on the LCD performance, the detectability using quarter-dose denoised outputs was inferior to that obtained using normal-dose CT scans. 
\end{abstract}

\begin{IEEEkeywords}
Deep Learning, Denoising, Image Quality, Task-based Performance
\end{IEEEkeywords}
\vspace{-1.0em}
\section{Introduction}
The diagnostic image quality of outputs obtained by applying denoising algorithms to low-dose CT images—compared to that exhibited by their normal-dose CT counterparts—remains unclear for clinical tasks such as lesions and nodule detection. This study investigates the image quality of denoised CT images using task-agnostic metrics like PSNR, SSIM, and clinical task-based performance for low contrast detectability (LCD). 

\vspace{-0.5em}
\section{Materials and Methods}
\subsection{Denoisers}
We built a library of non-linear CT image denoisers that can be classified into two groups, namely, (a) conventional denoisers and (b) Deep Learning (DL) denoisers. Our conventional denoisers' group consisted of Bilateral filtering \cite{bilateral_notes}, Total Variation (TV) \cite{chambolle2011first}, and Block Matching \& $3$D filtering (BM$3$D) \cite{bm3d_paper}. Our deep learning denoisers' group consisted of a $3$-layered Convolutional Neural Network(CNN3), $10$-layered Residual Encoder-Decoder CNN (REDCNN) \cite{red_net_orig}, Generative Adversarial Networks (GAN) \cite{goodfellow_GAN}, $17$-layered Feed-forward
denoising CNN (DnCNN) \cite{zhang_DnCNN}, and $10$-layered Dilated U-shaped DnCNN (UNeT) \cite{unet_git}. \blockcomment{Our GAN model comprised $20$-layered ResNet \cite{srgan_paper} as its generator network and the $10$-layered DnCNN as its discriminator network.}

The DL denoisers were trained using normal- and quarter-dose CT image pairs from six patients (L$096$, L$143$, L$192$, L$291$, L$286$, L$333$) provided in the 
2016 Low-Dose CT (LDCT) Grand Challenge database \cite{ldct_data}. The training set comprises $1560$ CT images of size $512 \times 512$. The CT images were reconstructed at $3$ mm slice thickness using the Filtered BackProjection (FBP) method with a sharp kernel (D$45$). A detailed description of training and tuning all the deep learning denoisers is provided in ref.~\cite{pkc_deep_ct}.

The conventional denoisers were optimized using $40$ CT quarter- and normal-dose CT pairs randomly sampled from the six patient data. Accordingly, the tuned parameters for the bilateral filtering - implemented using scikit-image \blockcomment{cite{skimg}} - consisted of window size for filtering as $7$, the standard deviation for gray value/color distance as $0.02$, the standard deviation for range distance as $5$. The tuned regularization parameter for the TV denoiser was $0.016$, and the $\sigma$ parameter for the BM$3$D denoiser was $0.034$.

\subsection{Image Quality Assessment}
The image quality of the denoised outputs was assessed using task-agnostic and task-based performance testing. For the task-agnostic assessment, we used visual perception and image fidelity-based metrics such as PSNR and SSIM. The normal and quarter-dose CT images of patient L$506$ from the LDCT database were used to estimate the PSNR and SSIM values.

For the task-based assessment, we analyzed the low contrast detectability of the four inserts in the simulated CT images of CCT$189$ phantom (in fig.\,\ref{img:lcd_of_denoisers}(a)). We simulated $200$ noisy $2$D scans from the CCT$189$ phantom at four dose levels, i.e., $25\%,\ 50\%,\ 75\%,\text{ \& } 100\%$ of normal (with the maximum incidence flux set as $0.85\times2.25\times10^{5}$). The exact process was repeated for simulating $100$ noisy scans from a $2$D uniform water phantom. For each of the four inserts in CCT$189$, one signal-present (SP) ROI was extracted from the CCT$189$ phantom image, and five signal-absent (SA) ROIs were extracted at the vicinity of the insert location from each of the uniform phantom images. As a result, $200$ SP \& $500$ SA ROIs for each low contrast insert were created to evaluate its corresponding detectability performance, particularly AUC, using Laguerre-Gauss Channelized Hotelling Observer (LG-CHO). $100$ SA \& SP ROI pairs were used to train the LG-CHO, and the remaining ROIs were used in the detectability testing. All the CT scans for the LCD test were simulated using a fan-beam CT projection model with Poisson noise and an FBP reconstruction algorithm provided in the Michigan Image Reconstruction Toolbox (\url{https://web.eecs.umich.edu/~fessler/code/}).
\vspace{-0.5em}
\subsection{Results}
\vspace{-2.0em}
\begin{table}[htbp]
    \caption{PSNR and SSIM values from different denoisers applied to quarter-dose patient test CT scans.}
    \begin{center}
    \label{tab:global_metric}
    \begin{NiceTabular}{c|c|c|c}
	\hline
	& \textbf{Denoisers} & \textbf{PSNR} (std) & \textbf{SSIM} (std) \\
	\hline
	 FBP \\quarter-dose &  - & $35.2$ ($\pm 2.5$) & $0.83$ ($\pm 0.07$)\\
    \hline
    \Block{4-1}<\rotate>{Conv-\\entional\\ denoisers}
    &     Bilateral        & $37.6$ ($\pm 3.0$) & $0.89$ ($\pm 0.06$) \\
    &      TV               & $37.6$ ($\pm 2.5$) & $0.91$ ($\pm 0.05$) \\
    &      BM$3$D           & $37.7$ ($\pm 2.4$) & $0.90$ ($\pm 0.04$) \\
    &                        &                    &                    \\
    \hline
    \Block{5-1}<\rotate>{Deep\\Learning\\ denoisers}
    &      CNN$3$           & $37.8$ ($\pm 2.9$) & $0.92$ ($\pm 0.04$) \\
    &      REDCNN           & $39.0$ ($\pm 2.9$) & $0.94$ ($\pm 0.03$) \\
    &      GAN              & $34.2$ ($\pm 2.3$) & $0.89$ ($\pm 0.03$) \\
    &      UNeT             & $34.7$ ($\pm 2.7$) & $0.85$ ($\pm 0.05$) \\
    &      DnCNN            & $38.6$ ($\pm 3.5$) & $0.94$ ($\pm 0.03$) \\
    \hline
	\end{NiceTabular}
    \end{center}
\end{table}

\begin{figure}[!htb]
\centering
\includegraphics[width=0.63\linewidth]{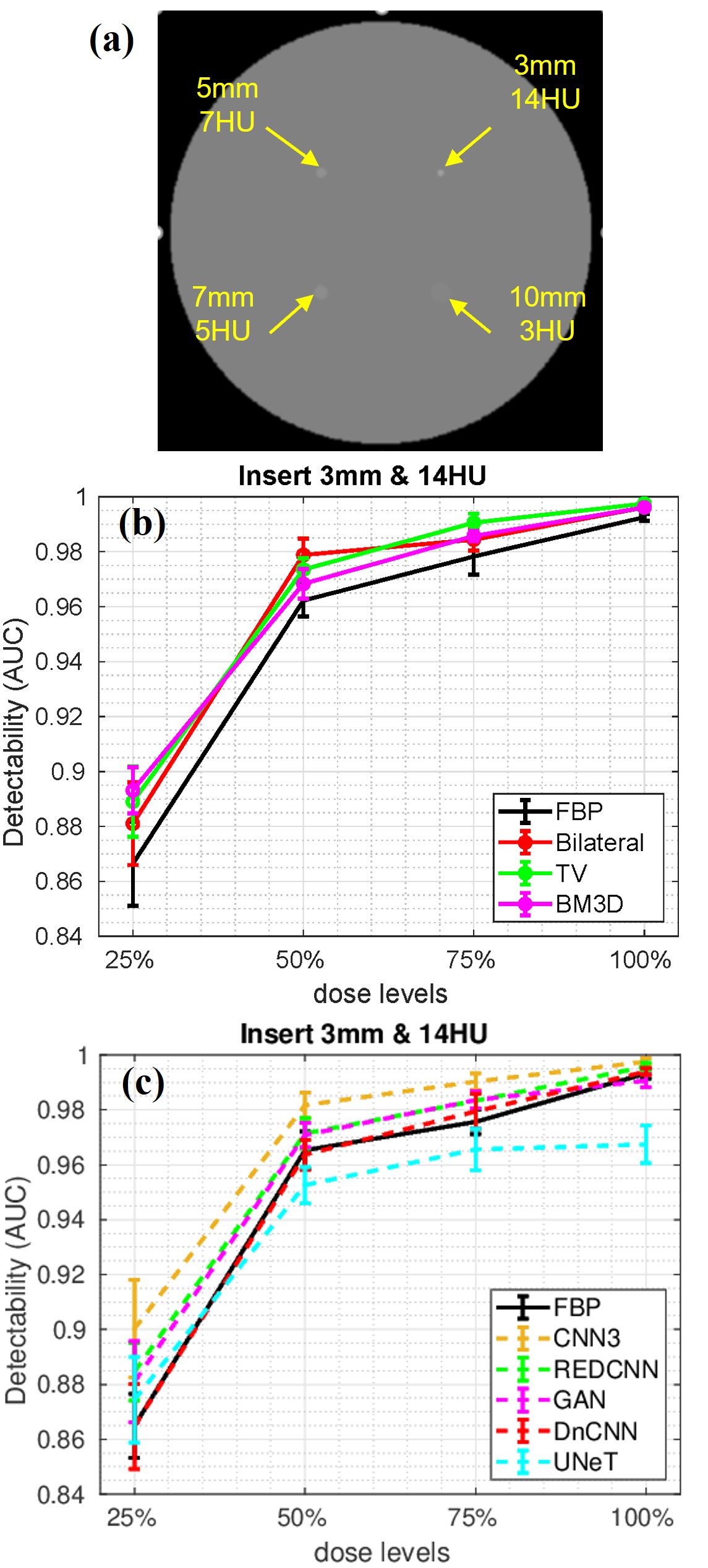}
\caption{(a) CCT$189$ low contrast body phantom for measuring the low contrast detectability. (b,c) LG-CHO-based detectability performance for the $3$mm/$14$ HU insert for different denoisers.}
\vspace{-1.0em}
\label{img:lcd_of_denoisers}
\end{figure}
\vspace{-1.0em}
PSNR and SSIM performance for all the denoisers is listed in Table \ref{tab:global_metric}. It is evident that all the denoisers (except UNeT and GAN) outperformed FBP quarter-dose by $2.4$ to $3.8$ dB in PSNR and $0.05$ to $0.11$ in SSIM. 
Likewise, figs.\,\ref{img:lcd_of_denoisers}(b,c) provide the AUC performance relative to the four dose levels on detecting the $3$ mm/ $14$ HU insert. For a given dose level, some denoisers improve while others decay the detectability performance. We did not find any substantial improvement in the denoisers’ LCD performance as the mean and standard deviation of the detectability AUC were similar between those obtained from the original and denoised quarter-dose CT images. Importantly, the detectability of quarter-dose denoised outputs was inferior to that obtained using normal-dose FBP scans. Similar detectability trends were observed for the other three inserts ($5$ mm/~$7$ HU, $7$ mm/~$5$ HU, $10$ mm/~$3$ HU). 
\vspace{-0.5em}
\subsection{Conclusion}
This study indicates that the gain in performance observed using an image fidelity-based (like the PSNR) or perceptual-based (like the SSIM) clinical task-agnostic metric may not correspondingly translate as a gain in detecting small-sized low-contrast signals (like lesions). The study employed simple uniform backgrounds with $2$D round disks as signals to be detected. In the future, we aim to validate further the performance trend observed in this study by analyzing a CAD algorithm's performance in finding lung nodules in denoised low-dose vs. normal-dose $3$D patient CT images. 

\vspace{-4.0mm}
\bibliographystyle{IEEEtran}
\bibliography{IEEEabrv}

\end{document}